\documentclass[a4paper,11pt,fleqn]{article}
\usepackage{pos}
\usepackage{slashed}

\title{First-principle evaluation of inclusive hadronic $\tau$~decays in QCD+QED}

\author*[a]{Matteo Di Carlo}
\author[b]{Simone Bacchio}
\author[c]{Erik B{\"a}ske}
\author[d,e]{Alessandro De Santis}
\author[f]{Antonio Evangelista}
\author[g]{Roberto Frezzotti}
\author[h]{Giuseppe Gagliardi}
\author[c]{Lukas Holan}
\author[h,i]{Vittorio Lubicz}
\author[g]{Lorenzo Maio}
\author[g]{Francesca Margari}
\author[c,j]{Agostino Patella}
\author[h]{Francesco Sanfilippo}
\author[h]{Silvano Simula}
\author[g]{Nazario Tantalo}

\affiliation[a]{Theoretical Physics Department, CERN,
1211 Geneva 23, Switzerland}

\affiliation[b]{
Computation-based Science and Technology Research Center, The Cyprus Institute, 20 Konstantinou
Kavafi Street, 2121 Nicosia, Cyprus
}

\affiliation[c]{
Humboldt Universit{\"a}t zu Berlin, Institut f\"ur Physik and IRIS Adlershof, Zum Gro{\ss}en Windkanal 6,\\ 12489 Berlin, Germany
}

\affiliation[d]{
Helmholtz-Institut Mainz, Johannes Gutenberg-Universit\"at Mainz, 55099 Mainz, Germany
}

\affiliation[e]{
GSI Helmholtz Centre for Heavy Ion Research, 64291 Darmstadt, Germany
}

\affiliation[f]{
Department of Physics, University of Cyprus, P.O. Box 20537, 1678 Nicosia, Cyprus
}

\affiliation[g]{
Dipartimento di Fisica and INFN, Universit\`a di Roma ``Tor Vergata'',
Via della Ricerca Scientifica 1, I-00133 Roma, Italy
}

\affiliation[h]{
Istituto Nazionale di Fisica Nucleare, Sezione di Roma Tre,
Via della Vasca Navale 84, I-00146 Rome, Italy
}

\affiliation[i]{
Dipartimento di Fisica, Universit\`a Roma Tre,
Via della Vasca Navale 84, I-00146 Rome, Italy
}

\affiliation[j]{
DESY, Platanenallee 6, D-15738 Zeuthen, Germany
}

\emailAdd{matteo.dicarlo@cern.ch}

\abstract{
We present a strategy to extend lattice calculations of inclusive hadronic $\tau$ decays from isosymmetric QCD to QCD+QED. 
The inclusive decay rate can be related to suitable Euclidean correlation functions, allowing for a first-principles evaluation of electromagnetic and isospin-breaking effects. 
Within the RM123 framework, radiative corrections are decomposed into leptonic, factorizable and non-factorizable contributions. 
We report preliminary results for the leptonic and factorizable terms in the electro-quenched approximation and discuss the remaining steps towards a complete calculation. 
This programme aims at a first-principles determination of inclusive $\tau$ decay rates with direct implications for the extraction of the CKM matrix element $|V_{us}|$.

\vspace{5mm}
\hfill \texttt{CERN-TH-2026-067}
}

\FullConference{The 42nd International Symposium on Lattice Field Theory (LATTICE2025)\\
2-8 November 2025\\
Tata Institute of Fundamental Research, Mumbai, India\\}

\usepackage[dvipsnames]{xcolor}
\usepackage{tikz}
\usetikzlibrary{positioning,patterns}
\usetikzlibrary{decorations.pathmorphing}
\usetikzlibrary{decorations.markings}
\usetikzlibrary{decorations.text}
\usetikzlibrary{shapes.misc}
\usetikzlibrary{arrows.meta,arrows}
\tikzset{->-/.style={decoration={
markings,
mark=at position .6 with {\arrow{>}}},postaction={decorate}}}
\tikzset{->n-/.style={decoration={
markings,
mark=at position .6 with {\arrow[scale=.7]{>}}},postaction={decorate}}}

\newcommand{\ket}[1]{\ensuremath{| {#1} \rangle }}
\newcommand{\bra}[1]{\ensuremath{\langle {#1} |}}

\renewcommand{\vec}[1]{\boldsymbol{#1}}

\begin{document}
\maketitle

\section{Introduction}

Inclusive hadronic decays of the $\tau$ lepton provide an important avenue for a first-principles determination of CKM matrix elements. 
Recent lattice calculations in isosymmetric QCD (isoQCD) have shown that the inclusive decay rate can be reconstructed from Euclidean lattice correlators with controlled uncertainties in both the $ud$ and $us$ channels~\cite{Evangelista:2023fmt,ExtendedTwistedMass:2024myu}. These studies also provided the first application, in this context, of the Hansen-Lupo-Tantalo (HLT) spectral reconstruction strategy proposed in Ref.~\cite{Hansen:2019idp}. In particular, the lattice determination of the strange inclusive rate has confirmed a tension of about $3\sigma$ between the value of $|V_{us}|$ extracted from inclusive $\tau$ decays and that obtained from kaon (semi)leptonic decays~\cite{ExtendedTwistedMass:2024myu}. Since this result has been obtained fully non-perturbatively in isoQCD, the discrepancy can no longer be attributed to the use of the operator product expansion in previous studies. This motivates extending the study beyond the isosymmetric approximation, including isospin-breaking and electromagnetic effects from first principles.

The study of inclusive hadronic $\tau$ decays beyond the isospin-symmetric limit is also relevant in a broader phenomenological context. 
In particular, hadronic $\tau$ data can provide complementary information on the isovector component of the hadronic-vacuum-polarization contribution to the muon anomalous magnetic moment. 
Dedicated investigations in this direction are currently being pursued both within chiral perturbation theory~\cite{Colangelo:2025ivq} and in lattice QCD~\cite{Bruno:2018ono}, with a primary focus on the $\tau\to\pi\pi\nu_\tau$ channel. In the present work, however, we concentrate on the first-principles determination of the fully inclusive hadronic decay rate in QCD+QED in the $us$ channel.

Our long-term strategy proceeds along two complementary directions. 
The first consists of a direct computation of the relevant Euclidean correlation functions in full QCD+QED, using gauge ensembles with dynamical photons generated by the RC$^\star$ collaboration and $C^\star$ boundary conditions~\cite{Lucini:2015hfa}. 
The second strategy is based on a perturbative expansion in $\alpha_{\rm em}$ and $(m_u-m_d)$ around isosymmetric QCD, following the RM123 approach~\cite{deDivitiis:2013xla}, and will employ the same physical-point gauge ensembles with $N_f=2+1+1$ Wilson-clover twisted-mass quarks that were used in the ETMC calculation of inclusive $\tau$ decays in isoQCD. 
While activity is ongoing in the first direction, in this work we concentrate on the second strategy. 
In particular, we outline the non-perturbative framework that relates the inclusive decay rate to Euclidean correlation functions in QCD+QED, and discuss its practical implementation within the RM123 approach, together with preliminary results.

The starting point is the weak effective Hamiltonian
\begin{align}
\mathcal{H}_W(x)
=
\frac{G_F}{\sqrt{2}}\sqrt{S_\mathrm{EW}}\,
\mathcal{O}_W(x)
=
\frac{G_F}{\sqrt{2}}\sqrt{S_\mathrm{EW}}\,
\big[\bar \tau \gamma^\alpha(1-\gamma_5)\nu_\tau\big](x)\,
j_\alpha(x)
+\mathrm{h.c.},
\end{align}
where $j_\alpha(x) = \sum_{q_1,q_2}  V_{q_1q_2} \; \bar{q}_1(x) \gamma^\alpha(1-\gamma_5) \, q_2(x)$ is the weak hadronic current, including the appropriate CKM factors, $G_F$ is the Fermi constant extracted from the muon lifetime (see e.g.~Sec.~10.2.1 of the PDG Review~\cite{PDG}) and $S_\mathrm{EW}$ takes into account process-dependent  electroweak corrections. 
Working at $\mathcal{O}(G_F^2)$ and to all orders in QCD+QED, one can derive an expression for the inclusive decay rate of the process
$\tau \to X\nu_\tau$, where $X$ denotes any flavoured hadronic state possibly accompanied by real photons allowed by kinematics. 
In the rest frame of the decaying $\tau$ lepton, introducing the momenta
\begin{align}
p=(m_\tau,\vec 0),\qquad
p_\nu=(|\vec p_\nu|,\vec p_\nu),\qquad
p_X=p-p_\nu,
\end{align}
the inclusive decay rate can be written as
\begin{align}
\Gamma
=
\frac{G_F^2S_\mathrm{EW}}{4m_\tau}\,
|\mathcal A(m_\tau)|^2,
\label{eq:decay_rate}
\end{align}
where the squared amplitude is given by
\begin{align}
|\mathcal A(m_\tau)|^2
=
\frac{1}{2}
\sum_r
\bra{\tau(r,p)}
\mathcal O_W(0)\,
(2\pi)^4
\delta^{(4)}(\hat P-p)\,
\mathcal O_W(0)
\ket{\tau(r,p)}_{C}.
\label{eq:amplitude}
\end{align}
Here the sum runs over the $\tau$ spin polarizations, $\hat P$ denotes the full QCD+QED four-momentum operator, and only the connected part of the matrix element is understood.


\section{Inclusive decay rate from a Euclidean correlator}
\label{sec:rate}

In this section we show how the inclusive decay rate can be related to a suitable Euclidean lattice correlation function. 
To this end, we assume that infrared divergences arising at intermediate stages of the calculation are properly regulated by introducing an infrared cut-off $\varepsilon_\gamma$, for instance by working in a finite volume $(\varepsilon_\gamma=1/L)$ or by assigning a small mass to the photon $(\varepsilon_\gamma=m_\gamma)$. 
The physical decay rate is then recovered in the limit $\varepsilon_\gamma\to 0$.

Working at $\mathcal{O}(G_F^2)$, the neutrino behaves as a free particle once produced. Its contribution can therefore be treated by replacing the field $\nu_\tau(0)$ with the corresponding spinors and summing over its polarizations.
By contrast, when QED effects are included to all orders the $\tau$ lepton is subject to electromagnetic interactions.
The extraction of the squared amplitude $|\mathcal{A}(m_\tau)|^2$ therefore requires the study of suitable correlation functions in QCD+QED and the isolation of the external $\tau$ states from their large-time behaviour, which can be achieved in the presence of the infrared regulator.

To this end we first introduce the Euclidean two-point function
\begin{align}
C_\tau(t) = \int d^3x\; T\bra{0} \tau(x)\, \bar \tau(0) \ket{0}\;, 
\end{align}
which, at finite $\varepsilon_\gamma$, is dominated at large positive Euclidean time $t$ by the single-$\tau$ state.

We then consider the Euclidean correlator
\begin{flalign}
&
\mathcal{C}(t_+,t,t_-)=
\frac{1}{2}\,\int d^3x_+ \, d^3x_- \,  d^3x \ 
T\bra{0}
\tau(t_++t,\vec x_+)\,
\mathcal{O}_W(t,\vec x) \mathcal{O}_W(0) \,
\bar \tau(-t_-,\vec x_-)
\ket{0}_C
\;,
\label{eq:C_full}\\[10pt]
\nonumber
& \qquad \qquad
\begin{tikzpicture}[scale=1.8]
    \def\mypath{(1,0) to [bend right=-40] ++(1,0) to [bend right=-40] +(-1,0)}
    \draw [-, line width=.4mm,  fill=black!30] \mypath;
    \pattern[pattern color=black, pattern=north east lines] \mypath;
    \draw [->n-, line width=.4mm, Green] (1,0) to [bend right=100] +(1,0);
    \draw [->-, line width=.4mm, blue] (0,0) to +(1,0);
    \draw [->-, line width=.4mm, blue] (2,0) to +(1,0);
    \node [above] at (0,0) {$-t_-$};
    \node [above] at (0.95,0.04) {$0$};
    \node [above] at (2.05,0.04) {$t$};
    \node [above] at (3,0) {$t+t_+$};
    \end{tikzpicture}
\end{flalign}
from which we define the ratio
\begin{flalign}
\mathcal{R}(t) \equiv 
\frac{4m_\tau \, \mathrm{Tr}\big[{\mathcal{C}}(t_+,t,t_-)\big]}{
\mathrm{Tr}\big[{C}_\tau(t_-+t_+)]}\;.
\label{eq:ratio_corr}
\end{flalign}
In the limits $t_+\gg 0$, $t_-\gg 0$ and for $t>0$, at finite $\varepsilon_\gamma$ and neglecting subleading exponentials, the ratio admits the spectral representation
\begin{equation}
\mathcal{R}(t)
=
\int_0^\infty
\frac{d\omega}{2\pi}\,
e^{-\omega t}\,
|\mathcal{A}(\omega)|^2 ,
\label{eq:R_vs_A2}
\end{equation}
thus providing a direct connection between the inclusive squared amplitude and a Euclidean lattice observable.

If one is able to invert the relation in Eq.~\eqref{eq:R_vs_A2} and extract the spectral density $|\mathcal{A}(\omega)|^2$ from $\mathcal{R}(t)$, the physical decay rate is obtained by evaluating the spectral density at $\omega=m_\tau$. 
This can be achieved through the HLT spectral reconstruction strategy outlined in Ref.~\cite{Hansen:2019idp}, which provides a solution to this inverse problem. 
Introducing a smeared approximation of the Dirac delta function satisfying
\begin{align}
    \lim_{\sigma\to 0} \Delta_\sigma(\omega,\bar\omega) = 2\pi \, \delta(\omega-\bar\omega)\,,
\end{align}
and assuming that, at fixed smearing radius $\sigma$, it can be accurately approximated by a linear combination of decaying exponentials,
\begin{align}
    \Delta_\sigma(\omega,\bar\omega) = \sum_t \, g_t(\sigma,\bar\omega) \, e^{-\omega t}\,,
\end{align}
one can finally obtain the rate as a linear combination of the timeslices of the Euclidean correlator,
\begin{flalign}
   \Gamma &=  \frac{G_F^2S_\mathrm{EW}}{4m_\tau} \int_0^\infty d\omega \; \delta(\omega-m_\tau) \, |\mathcal{A}(\omega)|^2
   = \lim_{\sigma\to 0} \; \frac{G_F^2S_\mathrm{EW}}{4m_\tau} \sum_{t}\, g_t(\sigma,m_\tau) \, \mathcal{R}(t)\;.
\end{flalign}
The challenge in such reconstruction is the reliable and controlled determination of the coefficients $g_t(\sigma,\bar\omega)$, which will be briefly discussed in Sec.~\ref{sec:reco}.


\section{RM123 expansion of the QCD+QED correlator}

The relations derived in the previous section hold for a fully non-perturbative treatment of electromagnetic interactions in QCD+QED. 
They can also be implemented by expanding the Euclidean correlator in powers of the electromagnetic coupling and the up-down quark mass difference. 
In this work we adopt the RM123 strategy~\cite{deDivitiis:2013xla}, in which isospin-breaking effects are incorporated through a systematic expansion around isoQCD. 
Within this framework the correlator $\mathcal{R}(t)$, and therefore the decay amplitude, can be expanded as
\begin{align}
\mathcal R(t)=\mathcal R^{(0)}(t)+\delta \mathcal R(t),
\end{align}
where $\mathcal R^{(0)}(t)$ is the isoQCD contribution. Taking derivatives with respect to the quark and lepton electric charges naturally leads to a decomposition of the radiative correction into three terms,
\begin{align}
\mathcal R(t)
=
\big[\mathcal R(t)\big]_{\rm lep}
+
\big[\mathcal R(t)\big]_{\rm fact}
+
\big[\mathcal R(t)\big]_{\rm non\mbox{-}fact},
\label{eq:R-split}
\end{align}
corresponding respectively to purely leptonic corrections (photon insertions on the $\tau$ line), factorizable hadronic corrections (photon exchanges among quark lines only), and non-factorizable contributions involving photon exchange between the lepton and hadronic sectors. In this notation, we include the isoQCD contribution among the factorizable corrections.

An important feature of this decomposition is that each contribution is separately infrared finite, making Eq.~\eqref{eq:R-split} particularly suitable for lattice implementations.
An analogous separation holds at the level of the squared amplitude,
\begin{equation}
|\mathcal A(m_\tau)|^2
=
|\mathcal A(m_\tau)|^2_{\rm lep}
+
|\mathcal A(m_\tau)|^2_{\rm fact}
+
|\mathcal A(m_\tau)|^2_{\rm non\text{-}fact}.
\end{equation}

As we will discuss below, a further simplification arises when working in isoQCD or when employing the RM123 approach.
In this case, part of the radiative correction can be expressed in terms of simpler purely hadronic Euclidean correlation functions. 
This occurs whenever the lepton and hadronic sectors do not interact through photon exchange, so that the neutrino can be completely factorized and treated as a free particle. 
The integration over the neutrino phase space can then be carried out analytically, leading to modified kernels that depend only on the energy of the intermediate hadron-photon system, rather than on the total internal energy. 


\subsection{Factorizable and leptonic contributions}

The factorizable contribution corresponds to radiative corrections in which photons are exchanged only among quark lines, together with the corresponding quark-mass counterterm contributions. 
In this case the kinematic structure of the decay amplitude is identical to the isoQCD case, while isospin-breaking effects enter through a modification of the hadronic spectral densities. 
The squared amplitude can be written as
\begin{equation}
|\mathcal{A}(m_\tau)|_\mathrm{fact}^2 
=
\frac{m_\tau^3}{2\pi}
\int_0^\infty \frac{dE}{2\pi}\,
E^2\,
\Big[
\mathcal{K}_{\mathrm T}(E/m_\tau)\,\rho^{\mathrm{full}}_{\mathrm T}(E^2)
+
\mathcal{K}_{\mathrm L}(E/m_\tau)\,\rho^{\mathrm{full}}_{\mathrm L}(E^2)
\Big].
\label{eq:Afact-rho}
\end{equation}
Here $\rho_{\mathrm T,\mathrm L}^{\mathrm{full}}$ denote the transverse and longitudinal spectral densities in full QCD+QED. 
The kernels $\mathcal K_{\mathrm T,\mathrm L}$ coincide with those of the isosymmetric theory and are determined by the leptonic kinematics. 
Introducing the dimensionless variable $x=E/m_\tau$, one finds
\begin{align}
\mathcal K_{\mathrm L}(x)
&=
\frac{1}{x}(1-x^2)^2\,\theta(1-x),
\qquad
\mathcal K_{\mathrm T}(x)
=
(1+2x^2)\,\mathcal K_{\mathrm L}(x).
\end{align}

The spectral densities entering eq.~(\ref{eq:Afact-rho}) are directly related to Euclidean hadronic correlation functions at zero spatial momentum,
\begin{align}
H_{00}^{\mathrm{full}}(t,\vec 0)
&=
\int_0^\infty \frac{dE}{2\pi}\, e^{-Et}\,E^2\,\rho_{\mathrm L}^{\mathrm{full}}(E^2),
\\[2pt]
\frac{1}{3}\sum_i H_{ii}^{\mathrm{full}}(t,\vec 0)
&=
\int_0^\infty \frac{dE}{2\pi}\, e^{-Et}\,E^2\,\rho_{\mathrm T}^{\mathrm{full}}(E^2),
\end{align}
with
\begin{align}
H_{\alpha\beta}^{\mathrm{full}}(t,\vec 0)
=
\int d^3x\;
T\!\bra{0}
j_\alpha(t,\vec x)\,
j_\beta^\dagger(0)
\ket{0}_{\mathrm{QCD+QED}} = \int_0^\infty \frac{dE}{2\pi} \, \rho^\mathrm{full}_{\alpha\beta}(E,\vec{0})\, e^{-E t}\,,
\end{align}
and
\begin{align}
    \rho^\mathrm{full}_{\alpha\beta}(q)= (q_\alpha q_\beta- q^2 \, g_{\alpha\beta}) \, \rho^\mathrm{full}_\mathrm{T}(q^2) + q_\alpha q_\beta \, \rho^\mathrm{full}_\mathrm{L}(q^2)\,.
\end{align}
Therefore, the factorizable correction can be obtained from the same class of Euclidean two-point functions used in the isoQCD analysis, the only difference being that the correlators must now be computed in full QCD+QED.

\vspace{6pt}

The purely leptonic correction has a complementary structure. 
Here photons are attached only to the external $\tau$ propagators, so that the hadronic dynamics is unchanged. 
Consequently the spectral densities remain those of isoQCD, while isospin-breaking effects appear through modified kinematic kernels. 
One obtains
\begin{align}
|\mathcal{A}(m_\tau)|^2_\mathrm{lep}
=
\frac{e_\tau^2 m_\tau^3}{16\pi^3}
\int_0^\infty \frac{dE}{2\pi}\,
E^2\,
\Big[
\delta\mathcal{K}_{\mathrm T}(E/m_\tau)\,\rho_{\mathrm T}(E^2)
+
\delta\mathcal{K}_{\mathrm L}(E/m_\tau)\,\rho_{\mathrm L}(E^2)
\Big].
\end{align}
The kernels $\delta\mathcal K_{\mathrm T,\mathrm L}$ arise from the sum of virtual and real photon emission from the lepton line. 
In terms of $x=E/m_\tau$ they read%
\footnote{The definition of the leptonic kernels depend on the definition of the QCD+QED $\tau$ wave function renormalization,~$Z_\tau$. Our choice here corresponds to $Z_\tau$ defined in the $W$-mass scheme~\cite{Sirlin:1981ie} at the renormalization scale $\mu=M_W$.}
\begin{align}
\delta \mathcal K_{\mathrm T}(x)
&=
\frac{1}{x}
\left\{
\left[
-\frac94+\frac12\log\frac{m_\tau^2}{m_W^2}
\right]
(1-x^2)^2(1+2x^2)
+
g_{\mathrm T}(x^2)
\right\}
\theta(1-x),
\\[2pt]
\delta \mathcal K_{\mathrm L}(x)
&=
\frac{1}{x}
\left\{
\left[
-\frac94+\frac12\log\frac{m_\tau^2}{m_W^2}
\right]
(1-x^2)^2
+
g_{\mathrm L}(x^2)
\right\}
\theta(1-x),
\end{align}
where
\begin{align}
g_{\mathrm L}(s)
&=
\frac{1}{12}
\Big[
(1-s)(61-65s-2s^2)
-6s^2\log s
-24(1-s)^2\log(1-s)
\Big],
\\[2pt]
g_{\mathrm T}(s)
&=
\frac{1}{12}
\Big[
(1-s)(61+37s-20s^2)
+6s^2(3+10s)\log s
-24(1+2s)(1-s)^2\log(1-s)
\Big].
\nonumber
\end{align}
Thus the leptonic correction can be reconstructed from the same Euclidean hadronic correlators entering the isoQCD analysis, the only modification being the replacement of the tree-level kernels with the radiatively corrected ones.

\vspace{6pt}

Finally, introducing a smeared version of the kernels and approximating them as linear combinations of decaying exponentials leads to reconstructed spectral densities of the form
\begin{flalign}
    |\mathcal{A}(m_\tau)|_\mathrm{fact}^2 
    &= \, \lim_{\sigma\to 0}\; 
    \sum_t \Big[ 
    g^\mathrm{(T)}_t(\sigma,m_\tau) \, H_{00}^\mathrm{full}(t,\vec 0) + 
    \frac{1}{3}\, g^\mathrm{(L)}_t(\sigma,m_\tau) \,  \sum_i H_{ii}^\mathrm{full}(t,\vec 0)
    \Big] \;,\\
    |\mathcal{A}(m_\tau)|^2_\mathrm{lep} & =  \, \lim_{\sigma\to 0}\; 
    \sum_t \Big[ 
    \delta g^\mathrm{(T)}_t(\sigma,m_\tau) \, H_{00}(t,\vec 0) + 
    \frac{1}{3}\, \delta g^\mathrm{(L)}_t(\sigma,m_\tau) \,  \sum_i H_{ii}(t,\vec 0)
    \Big] \;.
\end{flalign}
%


\section{Spectral reconstruction and present status}
\label{sec:reco}

The extraction of the physical amplitude from Euclidean data is an inverse problem and requires a suitable reconstruction strategy. In this work we rely on the HLT method~\cite{Hansen:2019idp}, in which the target kernel is approximated by a finite linear combination of decaying exponentials. The corresponding coefficients are determined by minimizing a functional that balances the accuracy of the kernel reconstruction against the statistical uncertainty of the result. 
At fixed smearing $\sigma$ and for a given target smeared  kernel $\mathcal{K}_\sigma(\omega,m_\tau)$, we minimize the functional
\begin{equation}
    W[\vec g,\lambda] = (1-\lambda) \, A[\vec g] + \lambda \, B[\vec g]\,,
    \label{eq:functional_W}
\end{equation}
where 
\begin{equation}
    A[\vec g] = \frac{1}{N_A} \, \int_{E_0}^\infty d\omega \, \big| \mathcal{K}_\sigma(\omega,m_\tau) - \sum_t g_t(\sigma,m_\tau) \, e^{-\omega t} \big|^2 \,, \quad N_A = \int_{E_0}^\infty d\omega \, \big| \mathcal{K}_\sigma(\omega,m_\tau) \big|^2\,,
\end{equation}
quantifies the deviation between the target and reconstructed kernels (with the energy $E_0$ chosen outside the support of the spectral density), while 
\begin{align}
    B[\vec g] = \sum_{t,s} g_t \,\mathbb{C}_{ts} \, g_s\,,
\end{align}
accounts for the statistical uncertainty of the reconstructed spectral density through the covariance matrix $\mathbb{C}$ of the Euclidean correlator. The free parameter $\lambda$ can then be varied and optimized through a stability analysis. For a complete and detailed description, we refer the reader to the Supplementary Material of Ref.~\cite{ExtendedTwistedMassCollaborationETMC:2022sta}.

While the details of the implementation of Euclidean lattice correlators including isospin-breaking corrections within the RM123 approach are deferred to a future publication, 
we present here preliminary results for the leptonic and factorizable corrections obtained on the D96 ensemble with Twisted-Mass fermions ($a \simeq 0.0569$~fm and $L \simeq 5.46$~fm) in the electro-quenched approximation, where electromagnetic effects involving sea quarks are currently neglected.%
\footnote{The implementation of factorizable corrections closely parallels that of the leading isospin-breaking effects to the hadronic-vacuum-polarization contribution to the muon $g{-}2$. The effort of the ETM Collaboration in this direction is described in Ref.~\cite{Evangelista:2025dzi}.} 
Data from additional ensembles are already available, and a comprehensive analysis is ongoing, including results at further values of the smearing parameter $\sigma$, different choices of the norm functional $A[\vec g]$, as well as for the alternative Osterwalder--Seiler lattice regularization for the quark fields.

\begin{figure}[hbt!]
    \centering
    \includegraphics[width=.48\textwidth]{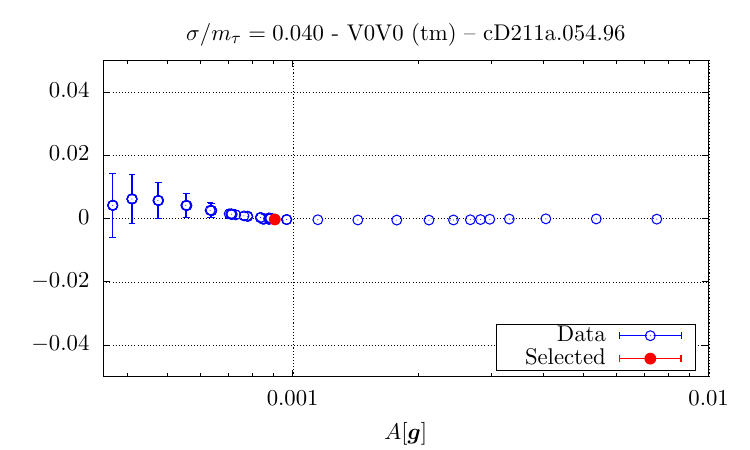}
    \includegraphics[width=.48\textwidth]{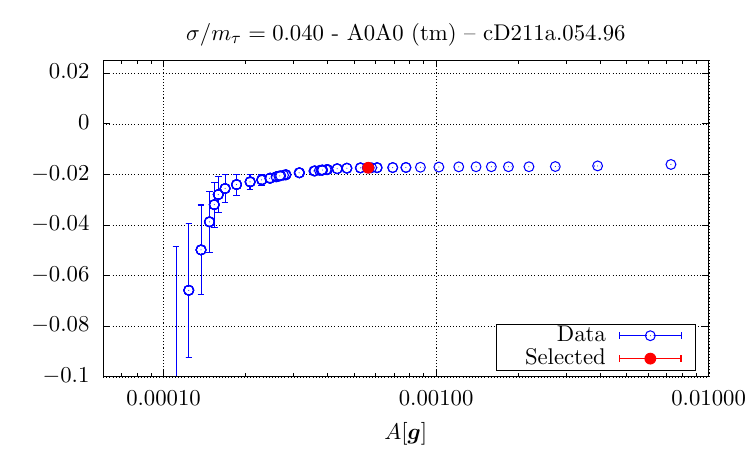}\\
    \includegraphics[width=.48\textwidth]{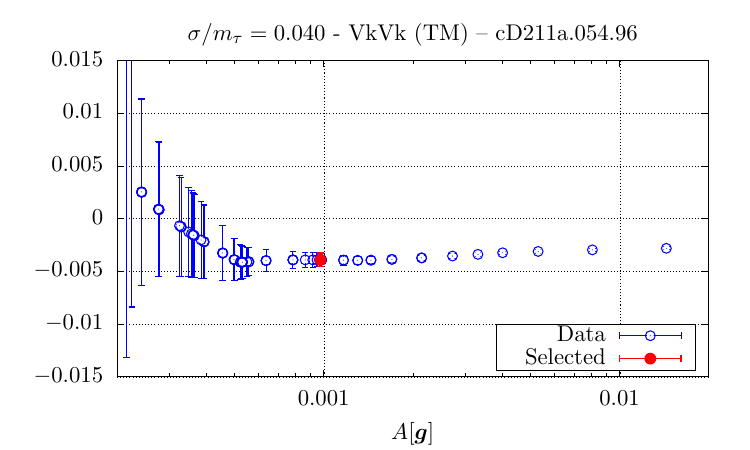}
    \includegraphics[width=.48\textwidth]{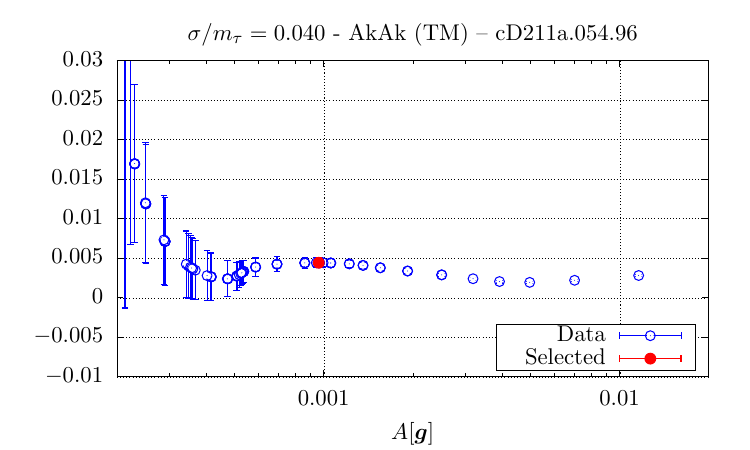}
    \caption{Preliminary reconstruction of the contribution of \emph{factorizable} corrections to the decay rate at $\sigma/m_\tau = 0.040$, obtained through a variation of the parameter $\lambda$. 
    \emph{Top panels}: reconstructed longitudinal spectral density from the temporal components of the vector (left) and axial (right) currents.
    \emph{Bottom panels}: reconstructed transverse spectral density from the spatial components of the vector (left) and axial (right) currents.
    }
    \label{fig:reco_fact}
\end{figure}
\begin{figure}[htb!]
    \centering
    \includegraphics[width=.48\textwidth]{figs/plot_V0_cD211a.054.96_tm_sigma0.040.pdf}
    \includegraphics[width=.48\textwidth]{figs/plot_A0_cD211a.054.96_tm_sigma0.040.pdf}\\
    \includegraphics[width=.48\textwidth]{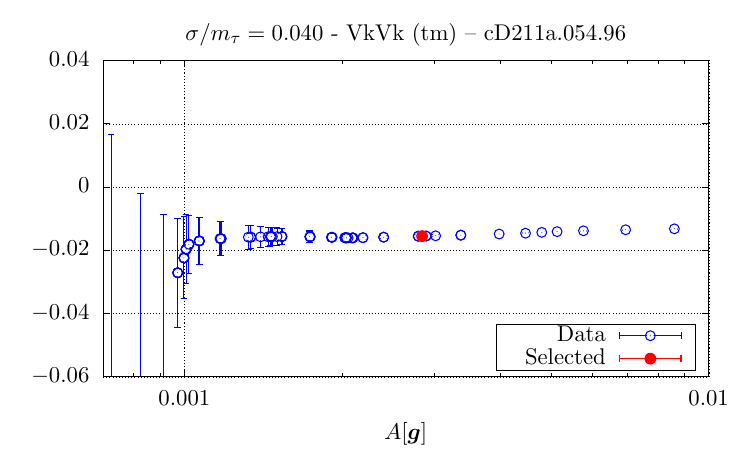}
    \includegraphics[width=.48\textwidth]{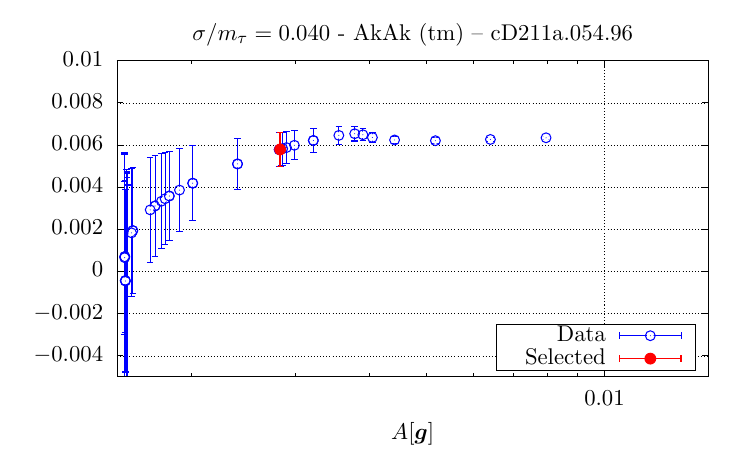}
    \caption{Preliminary reconstruction of the contribution of \emph{leptonic} corrections to the decay rate at $\sigma/m_\tau = 0.040$, obtained through a variation of the parameter $\lambda$. 
    \emph{Top panels}: reconstructed longitudinal spectral density from the temporal components of the vector (left) and axial (right) currents.
    \emph{Bottom panels}: reconstructed transverse spectral density from the spatial components of the vector (left) and axial (right) currents.
    }
    \label{fig:reco_lept}
\end{figure}

The corresponding reconstructed smeared spectral densities contributing to the factorizable and leptonic corrections to the inclusive decay rate, obtained for several values of the parameter $\lambda$ entering the functional in Eq.~\eqref{eq:functional_W}, are shown in Figs.~\ref{fig:reco_fact} and~\ref{fig:reco_lept}, respectively. 
In each figure, the red filled point represents the value of the reconstructed smeared spectral density at $\sigma/m_\tau = 0.040$ corresponding to the optimal choice $\lambda^*$, which lies in a region where the predicted spectral density remains stable within statistical uncertainty under variations of $\lambda$. 
As noted above, these results are preliminary, and a more robust analysis will include additional choices of kernel functions and functional norms in order to further constrain the reconstruction. 
It is nevertheless noteworthy that the reconstruction quality, as quantified by the value of $A[\vec g]$ at the optimal $\lambda^*$, is comparable to that obtained in the isoQCD study of Ref.~\cite{ExtendedTwistedMass:2024myu} for similar values of $\sigma$.


\section{Next steps: non-factorizable corrections and non-perturbative renormalization}

A class of isospin-breaking effects that has not been discussed so far is provided by the non-factorizable corrections, in which a photon is exchanged between the $\tau$ lepton and the hadronic system. 
In this case the hadronic spectral density cannot be factorized from the leptonic contribution, and the neutrino cannot be integrated out in a straightforward way. 
As a consequence, one must deal with a more involved Euclidean correlation function, to be combined with the smeared Dirac delta kernel introduced in Sec.~\ref{sec:rate}.
Alternatively, if one wishes to define reconstruction kernels that depend only on the energy of the hadron-photon system, the spectral reconstruction must be repeated for each value of the neutrino three-momentum $\vec p_\nu$.

The non-factorizable Euclidean correlation function can be written as
\begin{flalign}
    \big[\mathcal{R}(t)\big]_{\text{non-fact}} = e_\tau \; \frac{1}{L^3}\, \mathrm{Re} \int d^3 z\int d^3x \; \big[\, S_\nu(x, z)\, \big]_{ij} \; \big[\mathcal{K}_{ji}(x,z) \big]
    \Big|_{t=t_x-t_z}
    \;,
\end{flalign}
with
\begin{flalign}
    & \big[\mathcal{K}_{ji}(x,z) \big]
    =
    \int d^4y\; \Big[\gamma^\beta_L\, \mathcal{L}^\mu(y-z) \, \gamma^\alpha_L\Big]_{ji}\; T\langle 0 |  j^\alpha(x) \, j^\mu_\mathrm{em,H}(y) \, j^\beta(z)^\dagger |0 \rangle \;,\\[2pt]
    & \mathcal{L}_\mu(y-z) = 
    -
    \int \frac{d^4 k}{(2\pi)^4} \; S_\gamma(k) \, e^{ik\cdot (y-z)} \, \Big[S_\tau({p}-k) \gamma_\mu 
    (m_\tau-i\slashed{p})
    \Big]\;,
\end{flalign}
where $S_\nu(x,z)$, $S_\tau(p-k)$ and $S_\gamma(k)$ denote the neutrino, $\tau$-lepton and photon propagators, respectively, and $j^\mu_{\mathrm{em,H}}$ is the hadronic electromagnetic current.
The practical numerical implementation of this correlation function is currently under investigation, in particular regarding different strategies for treating the photon propagator, including both analytical representations and stochastic approaches.

Besides the treatment of non-factorizable corrections, another essential ingredient for a complete QCD+QED calculation is the renormalization of the weak effective Hamiltonian.
Once electromagnetic corrections are included, additional ultraviolet divergences arise and a consistent matching between the Standard Model, the continuum effective theory and the lattice regularization becomes necessary. 
On the lattice, fermion discretizations that break chiral symmetry, as in the present setup, induce operator mixings that are absent in the continuum and therefore require a dedicated non-perturbative renormalization programme. 
Similarly to the case of leptonic decays of pseudoscalar mesons studied in Ref.~\cite{DiCarlo:2019thl}, it can be shown that, upon averaging over the lepton Wilson parameter, the mixing pattern reduces to only two operators, corresponding to the mixing between hadronic currents of the type $(V{-}A)$ and $(V{+}A)$. Two complementary directions are currently being pursued: one based on lattice momentum-subtraction schemes, and the other employing the gradient flow.

The resolution of these challenges will pave the way for a first-principle non-perturbative determination of the inclusive decay rate in QCD+QED.


\section{Conclusions}

We have presented a strategy to extend the lattice study of inclusive hadronic $\tau$ decays from isosymmetric QCD to QCD+QED. The method is based on relating the inclusive decay rate to a Euclidean correlation function and reconstructing the physical amplitude through the HLT approach.

Within the RM123 framework, the radiative corrections naturally split into leptonic, factorizable and non-factorizable contributions. We have shown that the first two can be reformulated in terms of Euclidean hadronic correlators and suitable reconstruction kernels, and we have presented preliminary results for these contributions in the electro-quenched approximation. The observed reconstruction quality is encouraging and supports the feasibility of the programme.

The treatment of non-factorizable corrections and the non-perturbative renormalization of the weak Hamiltonian in QCD+QED remain the main open steps towards a complete calculation. Their implementation will be essential for achieving a fully first-principles determination of inclusive hadronic $\tau$ decay rates beyond isoQCD, with direct impact on the extraction of $|V_{us}|$.

\acknowledgments
M.D.C. has received funding from the European Union’s Horizon Europe research and innovation programme under the Marie Sk\l{}odowska-Curie grant agreement No.\ 101108006.
A.E. and S.B. acknowledge support from EXCELLENCE/0524/0017 (MuonHVP) and EXCELLENCE/0524/0459 (IMAGE-N), co-financed by the European Regional Development Fund and the Republic of Cyprus via the Research and Innovation Foundation under the Cohesion Policy Pro-
gramme ``THALIA2021–2027''.
A.E., R.F., G.G., L.M., F.M. and N.T. are supported by the Italian Ministry of University and Research (MUR) under the grant PNRR-M4C2-I1.1-PRIN 2022-PE2 Non-perturbative aspects of fundamental interactions, in the Standard Model and beyond F53D23001480006 funded by E.U.--NextGenerationEU.
F.S. is supported by ICSC--Centro Nazionale di Ricerca in High Performance Computing, Big Data and Quantum Computing, funded by European Union--NextGenerationEU and by Italian Ministry of University and Research (MUR) project FIS 00001556.

\vspace{-2mm}

\bibliographystyle{JHEP}
\setlength{\bibsep}{0pt}
\bibliography{refs}

\end{document}